# Gamma dose rate monitoring using a Silicon Photomultiplier-based plastic scintillation detector

Paolo Tancioni[1], Ulisse Gendotti[1]

1. Arktis Radiation Detectors Ltd
Raffelstrasse 11, 8045 Zürich, Switzerland

*Abstract-* Plastic scintillation detectors represent an efficient, cost-effective, customizable solution for assessment of gamma radiation fields in the context of nuclear security, environmental radiation survey, radiological and nuclear incidents response (RNe) and dosimetry for radiation protection purposes. Solid state light readout technologies such as Silicon Photomultiplier diodes (SiPM) are characterized by low operational voltage, mechanical robustness, compact dimensions, and fast pulse rise-time. Due to these intrinsic advantages, Silicon Photomultiplier diodes have been progressively replacing Photomultiplier Tube-based gamma detection technologies in a great part of these applications. In this experimental study, a gamma energy calibration and a dose rate calculation algorithm are discussed and implemented onto the SiPM-based Flat Panel Gamma detector (FPG) developed by Arktis Radiation Detectors Ltd (Zürich, Switzerland). The FPG detector has been deployed and studied in the framework of research projects on RNe technologies, involving its use on unmanned aerial and ground vehicle such as drones and robots. This work moved from the need for end users to have an accurate and reliable dose rate assessment when operating in the field, especially in the context of RNe response. Results are presented in terms of gamma dose rate accuracy obtained by laboratory measurements using radionuclide sources emitting in the 0.059 – 1.333 MeV gamma energy range, compared to theoretical dose rate calculations.

*Index Terms-* Silicon Photomultiplier (SiPM), Plastic Scintillation, Gamma radiation, Dose monitoring

## I. Introduction

Silicon Photomultiplier-based (SiPM) sensors can handle single-photon signals, allowing the detection of low intensity optical pulses with single photoelectron typical gain up to $10^6$. Moreover, SiPM operational voltage can be up to 30 times lower than the high voltage required by Photomultiplier Tubes (PMTs), while typically requiring around a factor 1000 less volume. These intrinsic advantages make the solid-state SiPM-based sensors considerably advantageous for applications requiring portable devices or dimension-limited designs. The possibility to directly couple SiPM sensors to bulk scintillators has been widely demonstrated and discussed over the last two decades, showing new possibilities in the field of gamma radiation detection with organic plastic scintillators [1-3]. In this article, a lightweight, portable Flat Panel Gamma (FPG) plastic scintillation detector with SiPM-based light readout produced by Arktis Radiation Detectors Ltd (Zürich, Switzerland) was used to develop a dose rate calculation algorithm exploiting its energy discrimination capabilities. A measurement campaign conducted at Arktis will be described, using common gamma-emitting radionuclide sources in the 0.059 – 1.333 MeV energy range. The generation of energy histograms for each isotope's emission spectrum will be described and the dose rate algorithm functioning will be outlined. In conclusion, the results on the dose rate calculation will be presented in terms of dose rate accuracy with respect to theoretical dose rate for each isotope.

## II. Flat Panel Gamma detector

The Flat Panel Gamma (FPG) consists of two slabs of plastic scintillator with different thickness. The front slab is thinner and is tailored for high efficiency in detecting low energy gamma rays such as for $^{241}$Am and $^{57}$Co. The rear slab is thicker, offering more stopping power for higher energy gamma rays such as the two emission features of $^{60}$Co. The plastic scintillator slabs having an active area of around 390 cm$^2$, are directly coupled to 24 SiPMs for light collection, each SiPM having a 6x6 mm$^2$ surface. The precise geometry of SiPMs' placement and optical coupling are proprietary and not disclosed by Arktis. A coincidence gate is implemented in the device to discriminate and cut out counts generated by dark current randomly triggering the SiPMs. Once the coincidence condition is satisfied, the electric signal generated by the SiPMs is summed up, digitized by the onboard signal-processing electronics, and distributed over a 192 bins histogram based on its Time-over-Threshold (i.e., the time during which the pulse is above a fixed electronic threshold).

The detector, shown in Figure 1, is powered and exchange data through an RJ45 connector, needing only one low-voltage cable to work.

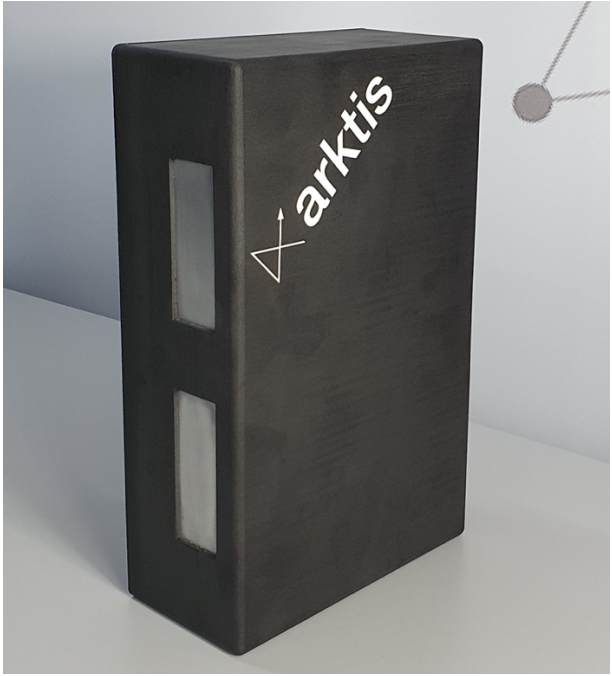

*Figure 1. The FPG detector by Arktis. Aluminum cooling plates are visible on the detector's side.*

With the goal of reducing noise due to temperature increase, aluminum cooling plates have been positioned on both the detector sides to enhance thermal exchange with the critical parts of the signal-processing electronics. The electrical and mechanical interface consists of another aluminum plate placed on the back surface of the detector, also serving as a cooling body. A section view of the detector is also shown in Figure 2.

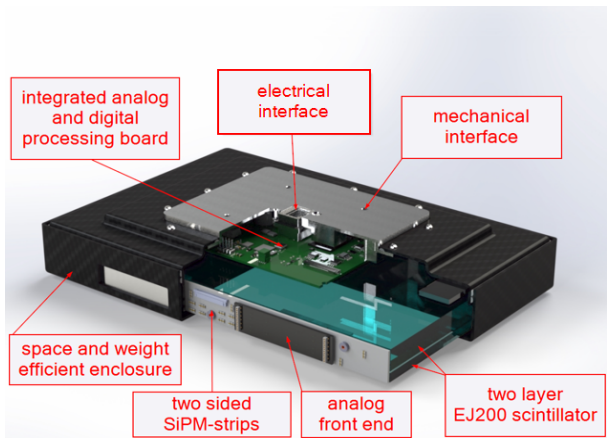

*Figure 2. Section and description of the detector's parts.*

## III. MEASUREMENT CAMPAIGN

A measurement campaign was performed in Arktis, consisting of a series of gamma irradiation tests with various radionuclide sources at different source-to-detector distances. In Figure 3, the FPG is being irradiated by a $^{241}$Am source. A list of the isotopes that were used in the measurement campaign, with their activities at day of test, main gamma emission energies and probabilities, is also shown.

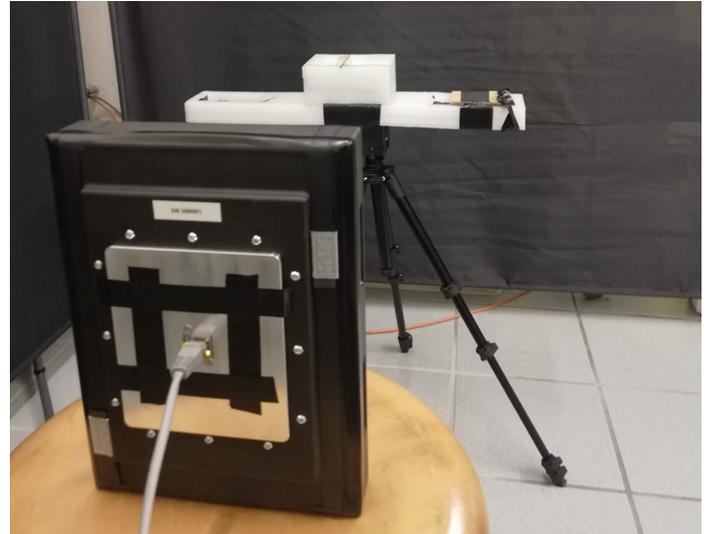

| Isotope | Activity [kBq] | Energy [keV] | Emission prob |
|---|---|---|---|
| Am-241 | 2343 | 59.54 | 0.359 |
| Co-57 | 898 | 122.10 | 0.856 |
|  |  | 136.50 | 0.107 |
| Ba-133 | 242 | 81.00 | 0.333 |
|  |  | 276.40 | 0.071 |
|  |  | 302.85 | 0.183 |
|  |  | 356.01 | 0.621 |
|  |  | 383.85 | 0.089 |
| Cs-137 | 3031 | 661.66 | 0.850 |
| Co-60 | 588 | 1173.23 | 0.999 |
|  |  | 1332.49 | 1.000 |

*Figure 3. The PFG detector during an irradiation measurement with a Am-241 source (top);*
*List of isotopes used in the measurement campaign with activities and gamma emission information (bottom).*

## IV. ENERGY CALIBRATION AND DOSE RATE ALGORITHM IMPLEMENTATION

In Figure 4, a superposition of the ToT histograms generated by the FPG for the different energy spectra in the experimental tests.

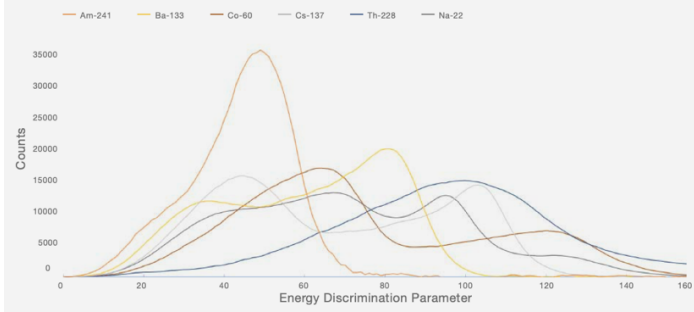

*Figure 4. Time-over-Threshold histograms for various gamma spectra gathered during the experimental tests with the FPG.*

As clearly visible, each irradiation measurements results in a different end point in the ToT histogram, allowing the discrimination of the gamma energy and thus of the radioisotope. In plastic scintillation detectors operating in the gamma energy range typical of these radionuclide sources, the main interaction mechanism is Compton Scattering.

The maximum of the Time-over-Threshold histogram corresponds to the Compton edge of the incident gamma. By correlation of the ToT histograms maxima to the incident energy for each isotope, an analytic fit of these points results in a logarithmic dependence, as shown in Figure 5.

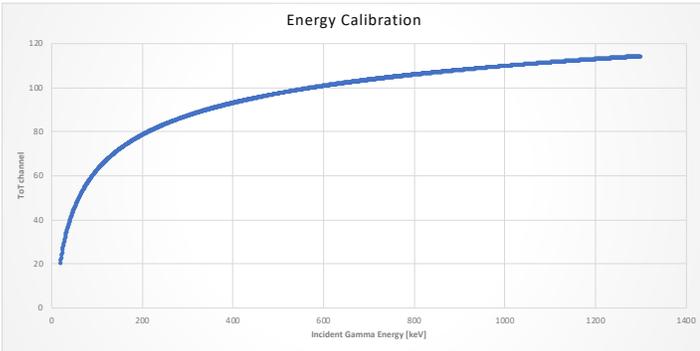

*Figure 5. Energy Calibration of the ToT histograms as a function of incident gamma energy Compton edge.*

Theoretical dose rate values on the detector surface were calculated for the listed radionuclide sources. During all the experimental tests the source-to-detector distance was kept at a distance large enough to ensure a homogeneous distribution of the dose rate field on the front surface.

When fitting the count rate ToT histogram to the theoretical calculated dose rate values, the best analytical fit of the multiplication function was given by an exponential. Therefore, the dose rate calibration function was written in an exponential form:

$$f(x_{ch}) = ae^{-bx_{ch}} + c$$

Where $x_{ch}$ is the ToT channel number and $a, b, c$ the calibration parameters. This function weights every channel of the histogram to fit the theoretical dose rate values. Once the experimental ToT histograms are gathered the dose rate is computed for each input ToT spectrum $i$:

$$D_i^{exp} = \sum_{ch} x_{ch} * f(x_{ch})$$

Three spectra were used with emission energies in the low, mid, and high energy range to provide a calibration function valid over the entire range of interest. $^{57}$Co, $^{137}$Cs and $^{60}$Co were chosen for this purpose, and the experimental ToT histogram is shown in Figure 6.

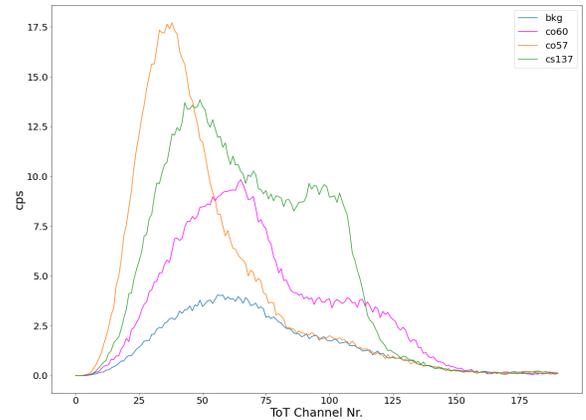

*Figure 6. Experimental ToT histograms for natural background, Co-57, Cs-137 and Co-60 sources.*

The least square method is used to have an indication of the difference the measured dose rates and the theoretical values, resulting in the loss function:

$$S = \sum_i (D_i^{exp} - D_i^{theo})^2$$

The algorithm minimizes this difference, by means of a simultaneous fit over the three input spectra to find the optimal set of parameters $a, b, c$.

## V. RESULTS AND CONCLUSION

The algorithm performance was tested by measuring dose rate values in the three energy regions while varying the source-to-detector distance from 50 cm to 2 m. The average accuracy of the dose rate calculation over the gamma energy range 0.059 – 1.333 MeV resulted to be of around 90%.

Am-241 59 KeV gamma emission was used to test the accuracy of the dose rate calculation at very low gamma energies,

showing the calculation consistency even in the low gamma energy region:

| Average Accuracy | Am-241 | Cs-137 | Co-60 |
|---|---|---|---|
| | 89% | 94% | 92% |

These results show that the Flat Panel Gamma, exploiting the characteristics of the Time-over-Threshold of the signals generated by the 24 Silicon Photomultipliers, can be a powerful dose rate survey instrument. Being lightweight, compact, and insensitive to mechanical shocks, this detection system can serve as a portable gamma detector for environmental radiological surveillance. Being powered by only one RJ45 cable and using only low voltage, this detector can easily be implemented onto unmanned vehicles such as robots or drones, allowing a fast, reliable, and accurate environmental dose rate assessment even in harsh environments.

This project has received funding from the European Union's Horizon 2020 research and innovation programme under Grant Agreement (GA) N° 786729. The views expressed in this paper reflect the views of the authors. The European Commission is not liable for its content and the use that may be made of the information contained herein.

## AUTHORS

**First Author** – Paolo Tancioni. Nuclear Engineer, Arktis Radiation Detectors Ltd.
**Second Author** – Ulisse Gendotti, PhD. CTO, Arktis Radiation Detectors Ltd.